\documentstyle[twocolumn,eqsecnum,aps]{revtex}

\newcommand{\half}{{\textstyle{\frac{1}{2}}}}
\newcommand{\rr}{{\rm I\mkern-3.5mu R}}   

\begin{document}
\draft

\title{Interfacial Energy and Fine Defect Structures \\
for Incoherent Films}

\author{Paolo Cermelli}
\address{Dipartimento di Matematica, Universit\`a
di Torino, Italy}
\author{Morton E. Gurtin}
\address{Department of Mathematical
Sciences,
Carnegie-Mellon University, Pittsburgh, PA, U.S.A.}
\author{Giovanni Leoni}
\address{Dipartimento di Scienze e Tecnologie Avanzate,
Universit\`a del Piemonte Orientale,
Italy}


\maketitle

\begin{abstract}
This note summarizes recent results \cite{cermelli} in which modern
techniques of
the calculus of variations are used to obtain qualitative features of 
film-substrate interfaces for a broad class of interfacial energies. In
particular, we show that
the existence of a critical thickness for incoherency and the
formation of interfacial dislocations depend strongly on
 the convexity and smoothness of the interfacial energy function.

 \end{abstract}


\section{Introduction}

        The structure of an interface  separating crystalline solids is
        strongly affected by the competition between 
the relaxation of the bulk crystals  to their respective equilibrium
        configurations, and the  tendency of the
        interface to mantain the
         exact matching (coherency) of the
          atoms of the two solids across the interface (cf. Leo \& 
          Hu \cite{leo1,leo2} and the review paper 
\cite{matthews}).
        When the stresses due to
          the deviation from equilibrium of the bulk material reach
           a suitable threshold,  interfacial
          dislocations appear that relax the bulk stresses and an
          extreme situation may be reached in which all
          regularity of the atomic bonding at the interface is lost.

We restrict attention to  a two-dimensional framework with corresponding
cartesian coordinates
 $(x,y)$ and basis $({\bf i}, {\bf j})$. We take the layer to be
infinite in the $x$-direction, denote by $h$ its height in the
$y$-direction, and
assume that $y
=0$ represents the interface between the layer and
the substrate, which we assume to be rigid.
Letting  ${\bf u}(x,y)$ denote
the displacement  of the layer, $\mbox{\boldmath $\epsilon$}=\half(\nabla{\bf
u}+\nabla{\bf u}^{\top})
$ the infinitesimal strain, and
$$
\mbox{\boldmath $\epsilon$}_{0}=\epsilon_0{\bf i}\otimes{\bf i},
$$
with $\epsilon_0>0$, the
{\it mismatch strain},
we consider the layer as elastic with  energy density
$$
w(\mbox{\boldmath $\epsilon$}-\mbox{\boldmath $\epsilon$}_0)=
\frac{E}{2(1+\nu)}\left\{
\frac{\nu}{1-2\nu}
[\mbox{tr}(\mbox{\boldmath $\epsilon$}-\mbox{\boldmath $\epsilon$}_0)]^{2}
+
\mbox{tr}[(\mbox{\boldmath $\epsilon$}-\mbox{\boldmath $\epsilon$}_0)^{2}]
\right\},
$$
with $E$ and $\nu$ Young's modulus and Poisson ratio, such that $E>0$ and
$-1<\nu<\half$.

We consider displacement fields
 ${\bf u}(x,y)$ that are  periodic in $x$,  ${\bf u}(1,y)={\bf
u}(0,y)+(\mbox{const.}){\bf i}
$ for $y\in[0,h]$;
therefore, modulo a rescaling,
 we may
restrict  attention to a cell of unit length $\Omega=[0,1]\times[0,h]$.
We  assume that the layer cannot separate from the substrate, so that
${\bf u}(x,0)\cdot{\bf j}=0$
 for $x\in[0,1]$. Thus, letting  $u(x)={\bf u}(x,0)\cdot {\bf i}$ denote
 the tangential displacement of the layer at the interface, 
we define the {\it incoherency strain} $\gamma(x)$ by
$$
\gamma(x)=\frac{du(x)}{dx},
$$
and refer to the interface as {\em coherent} if
$$
\gamma(x)=0
\qquad
x\in[0,1],
$$
and {\em incoherent} otherwise.

We assume that  the interfacial energy is an  even,  continuous function
$f(\gamma)$ of  the incoherency strain, and that
$f(0)=0$, while $f(\gamma)>0$
for $\gamma\ne 0$.
The total energy of the  system  is then given by
\begin{equation}
J({\bf u})=
\int_{\Omega}w(\mbox{\boldmath $\epsilon$}-\mbox{\boldmath $\epsilon$}_{0})\,dx\,dy
+
\int_{0}^{1}f(\gamma)\,dx.
\end{equation}

%


\section{Convexity of the interfacial energy density and existence of
smooth equilibrium states}

Equilibrium configurations of the layer correspond to states which minimize the
total energy: classically, such configurations correspond to
displacement fields
$\overline{\bf u}(x,y)$ such that
$$
J(\overline{\bf u})=\min_{{\bf u}\in W}J({\bf u})
$$
over  a suitable\footnote{Assuming that
$f$ satisfies the  growth condition $f(\gamma)\le C(1+|\gamma|^q)$,
with $C>0$ and $q\ge 1$,
we minimize $J$ on the space  $W$
of all functions ${\bf u}$ in the Sobolev space
$W^{1,2}(\Omega,{\rr}^{2})$, whose restriction to $\partial\Omega$ belong to
$ W^{1,q}(\partial\Omega,{\rr}^{2})$, which satisfy the periodicity
condition and are tangential to the interface.
} space $W$ of displacement fields on $\Omega$.

The chief problem here is that when
$f$ is not convex such minimizers $\overline{\bf u}$ may not exist. Even so,
one can obtain valuable physical insight by studying {\em minimizing
sequences}; that is, sequences $\{{\bf u}_n\}$ that tend to the infimum of the functional
$J({\bf u})$ in the sense that
$$
J({\bf u}_n)\to\inf_{{\bf u}\in W}J({\bf u}),
$$
as $n\to\infty$. In fact, as we shall see, such minimizing sequences help to
characterize  the microstructures resulting from various choices of the
interfacial energy $f(\gamma)$.

Central to this method  of analysis is the convex envelope $f^{**}(\gamma)$ of
$f(\gamma)$, which is the largest convex energy $f^{**}(\gamma)$ with
 $f^{**}(\gamma)\le f(\gamma)$  for all $\gamma$; precisely,  $f^{**}$ is the
supremum of all convex functions $g$ such that $g\le f$. Then $f$ is convex if
and only if $f=f^{**}$ (Fig. 1).

\par
 \vskip-.7cm
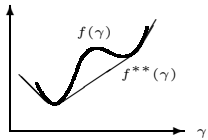
\begin{figure}
\setlength{\unitlength}{.042cm}
$$
 \begin{picture}(60,40)
 \put(-3,0){\vector(1,0){55}}
 \put(-3,0){\vector(0,1){40}}
 \thicklines
   \qbezier(5,15)(9,7)(12,9)
   \qbezier(12,9)(15,11)(18,20)
    \qbezier(18,20)(21,29)(28,25)
    \qbezier(28,25)(32.30769231,22.53846154)(35.30769231,24.53846154)
   \qbezier(35.30769231,24.53846154)(38.30769231,26.53846154)(40,33)
  \thinlines
 \put(12,8.4){\line(3,2){24}}
 \put(37.2,25.2){\line(3,5){6}}
 \put(8.9,9){\line(-1,1){9}}
   \put(56,-1){\tiny$\gamma$}
 \put(18,30){\tiny$f(\gamma)$}
\put(32,17){\tiny$f^{**}(\gamma)$}
 \end{picture}
 $$
 \caption{ The interfacial energy density (thick line)
  and its convex envelope (thin line). }
\end{figure}

The convexity of $f^{**}$ allows us to obtain a unique solution ${\bf u}^{**}$
of the "regularized problem"
$$
J({\bf u}^{**})=\inf_{{\bf u}\in W}
\left\{
\int_\Omega w(\mbox{\boldmath $\epsilon$}-\mbox{\boldmath $\epsilon$}_0)\,dx\,dy
+
\int_0^1f^{**}(
\gamma)\,dx
\right\}.
$$
In fact, this solution has the explicit form
\begin{equation}
{\bf u}^{**}(x,y):=\gamma^{**}\,x{\bf i}+\frac{\nu(\epsilon_0-\gamma^{**})}{
(1-\nu)}\,y{\bf j},
\label{tilde u-min}
\end{equation}
and corresponds to the {\em constant value $\gamma^{**}$ of the incoherency
strain $\gamma$} defined in Section IV below.

Consider the case in which $f$ is not convex at $\gamma^{**}$. Here (cf.
\cite{cermelli}) any minimizing sequence must satisfy\footnote{Convergence of
${\bf u}_n$ is here in the sense of
$W^{1,2}(\Omega,\rr^2)$.}
\begin{equation}
{\bf u}_n\to{\bf u}^{**}
\quad
\mbox{in}\ \Omega,
\quad
\mbox{and}
\quad
\int_0^1\gamma_n\,dx\to\gamma^{**},
\label{convergence}
\end{equation}
as $n\rightarrow\infty$.

Thus  minimizing sequences always converge in bulk to the homogeneous
deformation (\ref{tilde u-min}). As we shall see,
 minimizing sequences may not have a classical
limit at the interface,  but the ''generalized limit''\footnote{
The generalized or weak limit of a sequence 
can be thought of as the limit of its `'local integral averages''.}, however it be
visualized,
corresponds to a well defined average incoherency strain
$\gamma^{**}$.

Of course, ${\bf u}^{**}$ is  a candidate minimizer of the total energy,
but  since
\begin{equation}
J({\bf u}^{**})
=
\frac{Eh}{2(1-\nu^2)}
 (\gamma^{**}- \epsilon_0)^{2}
 +f(\gamma^{**}),
\label{added}
\end{equation}
then
\begin{eqnarray*}
(a)\ \ f(\gamma^{**})=f^{**}(\gamma^{**})
&\
\Leftrightarrow
&\
J({\bf u}^{**})=\inf_{{\bf u}\in W}J({\bf u}),
\\
(b)\ \ f(\gamma^{**})>f^{**}(\gamma^{**})
&\
\Leftrightarrow
&\
J({\bf u}^{**})>\inf_{{\bf u}\in W}J({\bf u}).
\end{eqnarray*}
Hence only when $f$ is convex at
 $\gamma^{**}$, does a minimum  exist for
the energy functional, at least in the classical sense.

\section{Minimizing sequences and interfacial microstructures}

Fix a thickness of the layer $h>0$,
let $\gamma^{**}$ be the average incoherency strain corresponding to the
infimum
of $J$, and assume that $\gamma^{**}\ne0$
and
$f(\gamma^{**})>f^{**}(\gamma^{**})$, so that no smooth equilibrium
state exists. In this case, physically meaningful results may be still
be obtained by inspection of the minimizing sequences $\{{\bf u}_{n}\}$.

By (\ref{convergence}) we may indeed restrict
attention to  the corresponding sequences of incoherency strains
$\gamma_n$ at the interface, which must, in turn, be  minimizing
sequences of the interfacial energy (cf. \cite{cermelli}):
\begin{equation}
\lim_{n\to\infty}\int_0^1 f(\gamma_{n}(x))\,dx=f^{**}(\gamma^{**}),
\label{minimizing-incoherency}
\end{equation}
with $f^{**}(\gamma^{**})<f(\gamma^{**})$.
We shall consider two  cases.
\par
\bigskip
\noindent
{\bf $\bullet$ Oscillating sequences (Fig. 2).}
Consider the situation in Figure 2(a), in which there is a $\lambda$,
$0<\lambda<1$, such that
\begin{eqnarray}
&&\gamma^{**}=\lambda \gamma_{a}+(1-\lambda)\gamma_{b},
\cr
&&
f^{**}(\gamma^{**})=\lambda f(\gamma_{a})+ (1-\lambda)f(\gamma_{b}).
\label{dacorogna-condition}
\end{eqnarray}
\par
 \vskip-.4cm
\begin{figure}
\setlength{\unitlength}{.042cm}
$$
 \begin{picture}(60,40)
 \put(0,0){\vector(1,0){40}}
 \put(0,0){\vector(0,1){50}}
   \put(41,0){\tiny$\gamma$}
  \qbezier(5,15)(8,15)(10,10)
  \qbezier(10,10)(20,50)(30,30)
 \qbezier(30,30)(32,40)(35,40)
\qbezier(10,10)(20,20)(30,30)
 \put(10,0){\dashbox(0,10){}}
\put(30,0){\dashbox(0,30){}}
\put(20,0){\dashbox(0,35){}}
\put(28,45){\tiny$f(\gamma)$}
\put(32,27){\tiny$f^{**}(\gamma)$}
\put(12,-5){\tiny$\gamma^{**}$}
 \put(3,-5){\tiny$\gamma_{a}$}
\put(30,-5){\tiny$\gamma_{b}$}
 \put(15,-13){\scriptsize(a)}
 \end{picture}
\quad\qquad
 \begin{picture}(50,40)
  \put(0,0){\vector(1,0){50}}
  \put(0,0){\vector(0,1){50}}
\put(4,20){\dashbox(0,20){}}
\put(10,20){\dashbox(0,20){}}
\put(14,20){\dashbox(0,20){}}
\put(20,20){\dashbox(0,20){}}
\put(24,20){\dashbox(0,20){}}
\put(30,20){\dashbox(0,20){}}
\put(34,20){\dashbox(0,20){}}
\put(40,0){\dashbox(0,40){}}
\put(0,20){\line(1,0){4}}
\put(4,40){\line(1,0){6}}
\put(10,20){\line(1,0){4}}
\put(14,40){\line(1,0){6}}
\put(20,20){\line(1,0){4}}
\put(24,40){\line(1,0){6}}
\put(30,20){\line(1,0){4}}
\put(34,40){\line(1,0){6}}
 \put(52,0){\tiny $x$}
 \put(20,45){\tiny $\gamma_n$}
 \put(-8,20){\tiny $\gamma_a$}
 \put(-8,38){\tiny $\gamma_b$}
 \put(40,-5){\tiny $1$}
\put(-3,-5){\tiny $0$}
 \put(18,-13){\scriptsize(b)}
 \end{picture}
 $$
 \par
 \vskip.5cm
 \caption{ (a) The interfacial energy density 
  and its convex envelope; (b) a typical element of a minimizing sequence 
  $\gamma_{n}$.}
\end{figure}
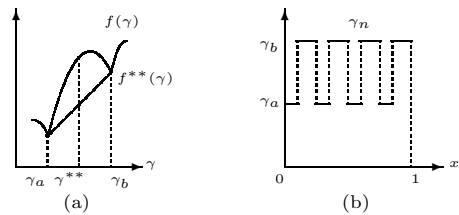
Then (\ref{minimizing-incoherency}) may be satisfied  by a minimizing
sequence  $\gamma_{n}$  oscillating between $
\gamma_{a}$ and $ \gamma_{b}$ on two subsets mixing finely as
$n\to\infty$. More precisely, we may take
\begin{equation}
\gamma_{n}(x)=
\left\{
\begin{array}{ll}
\gamma_{a}
&\quad x\in[\frac kn,\frac{k+\lambda}{n}
),
\\
\gamma_{b}
&\quad x\in[\frac{k+\lambda}{n},\frac {k+1}n),
\end{array}
\right.
\label{oscillating-sequence}
\end{equation}
with $k=0,\ldots,n-1$ (cf. Figure 2(b)). Thus, since for any $n$
$$
\int_0^1f(\gamma_n)\,dx=
\lambda f(\gamma_{a})+ (1-\lambda)f(\gamma_{b})
=f^{**}(\gamma^{**}),
$$
then (\ref{minimizing-incoherency}) is satisfied. Indeed,
the $\gamma_n$ are minimizers of the interfacial energy functional for fixed
$\gamma^{**}$, but their weak limit
$\gamma(x)\equiv\gamma^{**}$  is not.

Thus, in the limit as $n\to\infty$, the above sequence describes an interfacial
microstructure  whose corresponding  incoherency strain takes the values
${\gamma_a}$ and ${\gamma_b}$ on infinitesimal patches of length fractions
$\lambda$ and
$1-\lambda$ respectively.

\par
\bigskip
\noindent
{\bf $\bullet$ Concentrating sequences (Fig. 3).}
 Condition  (\ref{dacorogna-condition}) may not be satisfied in
the important case in which  $f$ is strictly concave (see also Fig.
6). We have
 $$
 f^{**}(\gamma)=m|\gamma|,
 \qquad
 \mbox{with}\
 m=\lim_{\gamma\rightarrow+\infty}\frac{ f(\gamma)}{\gamma}.
 $$

  \par
 \vskip-.4cm
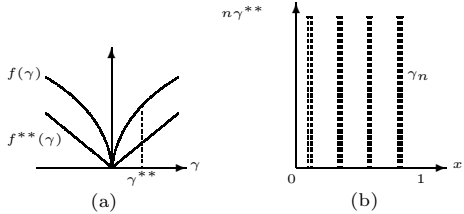
\begin{figure}
\setlength{\unitlength}{.04cm}
 $$
 \begin{picture}(60,45)
 \put(0,0){\vector(1,0){50}}
 \put(25,0){\vector(0,1){40}} 
  \put(51,0){\tiny$\gamma$}
  \put(-10,30){\tiny$f(\gamma)$}
 \put(-10,8){\tiny$f^{**}(\gamma)$}
\put(35,0){\dashbox(0,20.5){}}
\put(30,-5){\tiny$\gamma^{**}$}
  \qbezier(25,0)(27,18)(47,30)
  \qbezier(25,0)(23,18)(3,30)
 \qbezier(25,0)(36,9)(47,18)
 \qbezier(25,0)(14,9)(3,18)
 \put(18,-13){\scriptsize(a)}
 \end{picture}
\qquad \quad
 \begin{picture}(50,45)
  \put(0,0){\vector(1,0){50}}
  \put(0,0){\vector(0,1){55}}
\put(4,0){\dashbox(0,50){}}
\put(5,0){\dashbox(0,50){}}
\put(14,0){\dashbox(0,50){}}
\put(15,0){\dashbox(0,50){}}
\put(24,0){\dashbox(0,50){}}
\put(25,0){\dashbox(0,50){}}
\put(34,0){\dashbox(0,50){}}
\put(35,0){\dashbox(0,50){}}
\put(4,50){\line(1,0){1}}
\put(14,50){\line(1,0){1}}
\put(24,50){\line(1,0){1}}
\put(34,50){\line(1,0){1}}
 \put(52,0){\tiny $x$}
 \put(37,30){\tiny $\gamma_n$}
 \put(40,-5){\tiny $1$}
\put(-3,-5){\tiny $0$}
\put(-25,50){\tiny$n\gamma^{**}$}
 \put(18,-13){\scriptsize(b)}
 \end{picture}
 $$
 \par
 \vskip.5cm
 \caption{  (a) The interfacial energy density 
  and its convex envelope; (b) a typical element of a minimizing sequence 
  $\gamma_{n}$.}
\end{figure}

Fix $h>0$ as above and assume that $\gamma^{**}>0$. Then
(\ref{minimizing-incoherency}) can only be
satisfied by sequences $\gamma_n$ which become unbounded on smaller and smaller
sets, for instance
\begin{equation}
\gamma_{n}(x)=
\left\{
\begin{array}{ll}
\gamma^{**}n
&\quad x\in[\frac kn,\frac kn+\frac{1}{n^2}),
\\
0
&\quad x\in[\frac kn+\frac1{n^2},\frac {k+1}n),
\end{array}
\right.
\label{concentrating-sequence}
\end{equation}
for $k=0,\ldots,n-1
$. In this case, (\ref{minimizing-incoherency}) is satisfied since
$$
\int_0^1 f(\gamma_{n}(x))\,dx=
\frac{f(\gamma^{**}n)}{n}\to m\gamma^{**}
=f^{**}(\gamma^{**}).
$$
Note that, since the $\gamma_n$ tend to grow without bound on small sets,
the corresponding
displacements $u_n$  at the interface tend to develop microscopic jumps;
these may be identified with interfacial
dislocations.


\section{Critical thickness and smoothness of the convexified
interfacial energy density}

We now study the
behavior of the minimizers when  $h$ varies. We show that, when $f^{**}$
is smooth at zero,
the interface is incoherent (i.e., $\gamma^{**}\ne0$) for any $h>0$, while if
$f^{**}$ is non-smooth at zero, there exists a critical thickness $h_c$
such that for
$h<h_c$ the interface is coherent (so that $\gamma^{**}=0$).

Consider first a smooth $f^{**}$.  Let $j(\gamma)$ denote the right side of
$(\ref{added})$ when $\gamma^{**}$ is replaced by $\gamma$. Then
$j$  is smooth and convex and
$\gamma^{**}$ is the unique solution of
$j^\prime(\gamma)=0$. Since a direct calculation shows that $j^\prime(0)$ can
only vanish when $h=0$ then, for $h>0$, $\gamma^{**}\ne0$ and the interface is
incoherent.


Assume now that $f^{**}$, and thus $j$, is non-smooth at zero. Then the
condition
$j^\prime(0)=0$ must be replaced by  \ $0\in[j^\prime_-(0),j^\prime_+(0)]$, \
where $j^\prime_{\pm}$ are the left and right derivatives of $j$.
This condition is in general satisfied by  an interval of
values for the thickness  (cf. \cite{cermelli}), and thus the
equilibrium interface is coherent for all
$h\le h_c$, with $h_c$ a suitable critical thickness.

In general, the convexified interfacial energy $f^{**}$ may be
non-smooth at other values $\overline\gamma$  of the incoherency strain:
if this is the case,
we assume that  $f^{**}(\overline\gamma)=f(\overline\gamma)$, so
that a regular solution exists to the minimization problem.
 Proceeding as above, we see that the
condition which assures that $\gamma^{**}=\overline\gamma$ is that
$0\in[j^\prime_-(\overline\gamma),j^\prime_+(\overline\gamma)]$,
and this condition again determines an interval for the
thickness for which the incoherency strain remains fixed at the value
$\gamma(x)\equiv\gamma^{**}=\overline{\gamma}$: the film remains `glued' to
the  substrate.

\section{Discussion}

We now turn to the analysis of specific forms of the interfacial
energy density $f$ and discuss the existence of smooth
equilibrium configurations for the film,
the formation of microstructures at the interface, and the
critical thickness for incoherency.
Remark that the height of the film $h$ is now
allowed to vary, so that the
average incoherency strain $\gamma^{**}$ also varies accordingly.

\par
\bigskip
\noindent
{\bf $\bullet$ $f$ convex and smooth (Fig. 4(a)).}
Since in this case $f(\gamma)=f^{**}(\gamma)$, a regular solution exists
and is given by ${\bf u}^{**}$  in (\ref{tilde u-min}). In fact, the
restriction
of
${\bf u}^{**}$ to the interface, namely
$u(x)=\gamma^{**}x$, is a smooth minimizer of the interfacial energy
functional.
\par
\vskip-.8cm
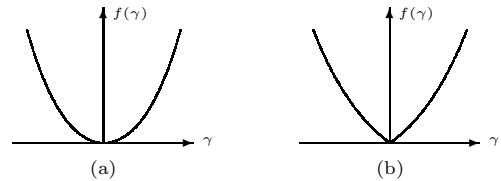
\begin{figure}
 \setlength{\unitlength}{.06cm}
 $$
 \begin{picture}(40,30)
 \put(0,0){\vector(1,0){40}}
 \put(20,0){\vector(0,1){30}}
 \put(42,0){\tiny$\gamma$}
 \put(22,28){\tiny$f(\gamma)$}
 \qbezier(20,0)(30,0)(37,25)
 \qbezier(3,25)(10,0)(20,0)
  \put(17,-7){\scriptsize (a)}
 \end{picture}
\qquad\qquad
 \begin{picture}(40,20)
 \put(0,0){\vector(1,0){40}}
 \put(20,0){\vector(0,1){30}}
 \put(42,0){\tiny$\gamma$}
 \put(22,28){\tiny$f(\gamma)$}
 \qbezier(20,0)(30,7)(37,25)
 \qbezier(3,25)(10,7)(20,0)
 \put(17,-7){\scriptsize (b)}
 \end{picture}
 $$
 \par
\vskip.3cm
 \caption{Interfacial energy density: (a) smooth and 
convex; (b) non-smooth 
but convex.}
\end{figure}
 Thus
the film is uniformly  incoherently strained with respect to the
substrate, 
but no fine structure appears.
Moreover, since $f^{**}(\gamma)$ is smooth, the interface relaxes to
incoherency for any
thickness of the layer.

\par
\bigskip
\noindent
{\bf $\bullet$ $f$ convex  but non-smooth at $\gamma=0$ (Fig. 4(b)),}
Again $f(\gamma)=f^{**}(\gamma)$, so that the homogeneous displacement
${\bf u}^{**}$ in (\ref{tilde u-min})
is a minimizer of $J$, and
no fine structure develops at the interface.

Now, since $f$ is non-differentiable at $\gamma=0$, there exists a critical
thickness for the transition to incoherency
such that  for $h\le h_{c}$ the interface is coherent, while
 for $h>h_{c}$
 the interface is uniformly strained with respect to the substrate.

This form of interfacial energy might be appropriate to describe
 ``glassy'' interfaces between crystals with large mismatch.

\par
\bigskip
\noindent
{\bf $\bullet$ $f$ nonconvex and nonsmooth  (Fig. 5).}
We assume here that $f$ has pointed minima at values
$\gamma_i\in\{\gamma_0=0,\pm
\gamma_1,
\pm\gamma_2,\dots\}$ of the incoherency strain, so that the convexified
interfacial
energy $f^{**}$ is piecewise linear in the intervals $(0,\gamma_1)$,
$(\gamma_1,\gamma_2)$,$\dots$, but non smooth at  $\gamma=\gamma_i$.
\par
\vskip-.4cm
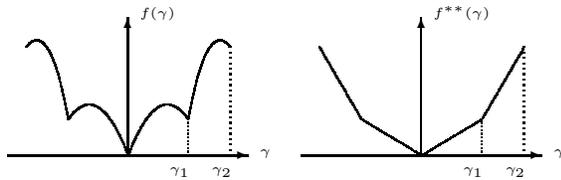
\begin{figure}
 \setlength{\unitlength}{.08cm}
$$
 \begin{picture}(40,20)
 \put(0,0){\vector(1,0){40}}
 \put(20,0){\vector(0,1){23}}
 \put(42,0){\tiny$\gamma$}
 \put(22,23){\tiny$f(\gamma)$}
 \put(27,-3){\tiny$\gamma_{1}$}
 \put(34,-3){\tiny$\gamma_{2}$}
 \bezier{10}(30,0)(30,3)(30,6)
 \bezier{20}(37,0)(37,9)(37,18)
 \qbezier(20,0)(25,13)(30,6)
 \qbezier(30,6)(33,23)(37,18)
 \qbezier(20,0)(15,13)(10,6)
 \qbezier(10,6)(7,23)(3,18)
 \end{picture}
 \qquad
 \begin{picture}(40,20)
 \put(0,0){\vector(1,0){40}}
 \put(20,0){\vector(0,1){23}}
 \put(42,0){\tiny$\gamma$}
 \put(22,23){\tiny$f^{**}(\gamma)$}
\put(27,-3){\tiny$\gamma_{1}$}
 \put(34,-3){\tiny$\gamma_{2}$}
 \bezier{10}(30,0)(30,3)(30,6)
 \bezier{20}(37,0)(37,9)(37,18)
 \qbezier(20,0)(25,3)(30,6)
 \qbezier(30,6)(33.5,12)(37,18)
 \qbezier(20,0)(15,3)(10,6)
 \qbezier(10,6)(6.5,12)(3,18)
 \end{picture}
 $$\par
\vskip.1cm
 \caption{Nonsmooth scalloped energy $f$ and its
 convex envelope $f^{**}$.}
\end{figure}
\par
\noindent(i) Existence: since $f(\gamma)=f^{**}(\gamma)$ only for
$\gamma=\gamma_i$, the homogeneous deformation ${\bf u}^{**}$ in
(\ref{tilde u-min}) is a minimizer only when  $\gamma^{**}=\gamma_i$.
When $\gamma^{**}=0$ the interface is coherent, while for $\gamma^{**}=\pm
\gamma_1, \dots$ the film is uniformly strained with respect to the substrate.
\par
\noindent
(ii) Critical thickness: since $f^{**}$ is non-smooth at $\gamma_i$, the
requirement that $\gamma^{**}=\gamma_i$ has the form $0\in
[j^\prime_-(\gamma_i),j^\prime_+(\gamma_i)]$.
Thus, there exists a whole interval for $h$
for which this condition is satisfied. More precisely, there exist critical
intervals
$[0,h_0]$, $\dots$, $[h_i^-,h_i^+]$ such that if
$h\in[h_i^-,h_i^+]$, then $\gamma^{**}=\gamma_i$. This means that the interface
remains  ``glued" to the substrate at this fixed incoherency strain for all
values of
the thickness in the critical interval.
\par
\noindent
(iii) Microstructures: for all other values of $h$ we have  $\gamma^{**}\ne
\gamma_i$,   and (\ref{tilde u-min}) is not a minimizer of $J$.
But since  condition (\ref{dacorogna-condition}) holds, the
minimizing sequences
have an oscillating character, and are finer and finer mixtures of patches
on which
$\gamma$ takes the values $\gamma_a=0$ and $\gamma_b=\gamma_1$ (if
$\gamma^{**}\in(0,\gamma_1)$), or
$\gamma_a=\gamma_1$ and
$\gamma_b=\gamma_2$ (if $\gamma^{**}\in
(\gamma_1,\gamma_2)$), and so on.

This behavior  is reminiscent of
"coincidence boundaries'', which are special incoherent interfaces between
crystals
with a large difference in lattice parameters.

\par
\bigskip
\noindent
{\bf $\bullet$ $f$  concave and non-smooth at $\gamma=0$  (Fig.
6).}
We have seen that in this case
$$
f^{**}(\gamma)=m|\gamma|,
$$
with  $f(\gamma)=f^{**}(\gamma)$ only for $\gamma=0$. Thus the only smooth
solution to the minimization problem corresponds to $\gamma^{**}=0$, a 
coherent
interface. For values of $h$ at which $\gamma^{**}\ne 0$ coherency is lost and
fine microstructures appear at the interface.
\par
\vskip-.5cm
\begin{figure}
  \setlength{\unitlength}{.08cm}
$$
 \begin{picture}(40,20)
 \put(0,0){\vector(1,0){40}}
 \put(20,0){\vector(0,1){23}}
 \put(42,0){\tiny$\gamma$}
 \put(22,21){\tiny$f(\gamma)$}
 \qbezier(20,0)(22,10)(37,15)
 \qbezier(20,0)(18,10)(3,15)
 \end{picture}
 \qquad
 \begin{picture}(40,20)
 \put(0,0){\vector(1,0){40}}
 \put(20,0){\vector(0,1){23}}
 \put(42,0){\tiny$\gamma$}
 \put(22,21){\tiny$f^{**}(\gamma)$}
 \qbezier(20,0)(28.5,4)(37,8)
 \qbezier(20,0)(11.5,4)(3,8)
 \end{picture}
 $$
 \caption{Nonsmooth strictly concave energy $f$ and its
 convex envelope $f^{**}$.}
\end{figure}
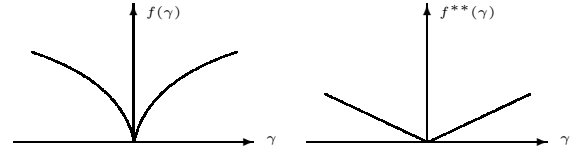
More precisely, observe first that
$f^{**}$ is non-smooth at zero, so that there exists a critical thickness
$h_c$ for
incoherency.
When $h>h_c$, since $\gamma^{**}\ne0$, the displacement field ${\bf u}^{**}$
in (\ref{tilde u-min})  is
 no longer a solution; the minimizing sequences are concentrating,
and energy is
minimized by allowing the incoherency strain to become infinitely large on
infinitesimal intervals: accordingly, the interfacial displacement
$u_{n}$ tends to
develop microscopic jumps, which we may interpret as interfacial
dislocations.

These energies might  describe small-misfit epitaxial layers.

\acknowledgements
M.G. was supported
by the U.S. Department of Energy and N.S.F.
P.C  and  G.L.
have been supported by the
M.U.R.S.T.  (``Modelli Matematici per la Scienza dei
Materiali'') and the
C.N.R.  (``Modelli e Metodi per la Matematica
e l'Ingegneria'') respectively.

\end{document}